\documentclass[twocolumn,showpacs,amsmath,amssymb,letterpaper]{revtex4}

\usepackage{graphicx}
\usepackage{dcolumn}
\usepackage{bm}
\usepackage{hyperref} 
\usepackage{multirow} 

\newcommand{\req}[1]{Eq.\,(\ref{eq:#1})}
\newcommand{\labeq}[1]{\label{eq:#1}}
\begin{document}
\title{Multistrange Particle Production and the Statistical Hadronization Model}%
\author{Michal Petr{\'a}{\v n}}
\author{Johann Rafelski}%
\affiliation{%
Department of Physics, University of Arizona, Tucson, Arizona 85721
}%

\date{December 9, 2009}

\begin{abstract}
We consider the chemical freeze-out of  $\Xi$, $\overline\Xi$ and $\phi$
multistrange hadrons within a Statistical Hadronization Model inspired 
approach. We study particle yields across a wide range 
of reaction energy and centrality  from  NA49 at SPS  
and STAR at RHIC experiments. We constrain the physical 
conditions present in the  fireball  source of strange hadrons,
and  anticipate results expected at LHC.
\end{abstract}


\maketitle

{\bf Introduction:}
We study multistrange hadron production in the context of the 
quark-gluon plasma (QGP) formation in relativistic heavy 
ion collisions~\cite{Timmins:2009vy}.   Given the relatively small
reaction cross sections of multistrange hadrons in hadron matter,  
the observed   yields   of  $\Xi(qss),$ $\overline\Xi,$ 
$\Omega(sss),$ $ \overline\Omega,$ 
$ \phi(s\bar s)$~\cite{Abelev:2008zk,Alt:2008iv,Alt:2008qm,Adams:2006ke,Speltz:2006,Abelev:2008ez}
are considered probes of the earliest stage of the  
QGP-fireball hadronization.

The yields of these
particles have been considered previously  within 
a global approach, see e.g.~\cite{Kuznetsova:2009wd}. Here
we show that it is possible to analyze multistrange hadron  yields alone. 
When this is done we  find that multistrange and non-strange 
hadrons share the same freeze-out condition. We will discuss the meaning
of this discovery below addressing dynamics of hadronization.
We also address the forthcoming LHC effort to measure
 multistrange hadron yields in high multiplicity 
$pp$~\cite{ZLM}, and soon after, in   A+A reactions.

QGP hadronic particle production yields are usually 
considered within the statistical hadronization 
model (SHM)~\cite{Letessier:2002gp,Torrieri:2004zz,Becattini:2009fv}.
SHM has been successful in describing (strange) hadron production in
heavy ion collisions for different colliding systems and 
energies. These results showing successful global fits of particle yields 
in the SHM framework inspired us to study multistrange hadron yields alone in this separate
analysis for the purpose of: i) establishing that SHM is appropriate for  describing
yields of these particles, ii) assessing if their yields are consistent with the 
established bulk matter properties of the QGP fireball, thus testing the single 
freeze-out hypothesis for particles with large and small hadron reaction cross sections,
and iii)  understanding better how the future LHC results may help arrive at
 a distinction between SHM model approaches.

{\bf SHM Models:} We begin by introducing the three principal SHM approaches:

{\bf a)} 
Taking the view that SHM has a limited theoretical foundation, one can
seek   {\it simplicity}   in an effort to obtain a qualitative 
description of the yields for all hadrons with just 
a small number of parameters. An additional attraction is that
this assumption leads to a model with  chemical equilibrium hadron yields 
is explored.  The main result of this approach is that
the hadronization in high energy heavy ion collisions at RHIC requires $T\ge 175$\,MeV,
and this high value is close to the lattice crossover temperature,  
between deconfined and hadron phase~\cite{Fodor:2009ax,BraunMunzinger:2008tz}. 

{\bf b)} In order to arrive at a precise description of the bulk properties, 
such as strangeness and entropy content of the 
hadron   fireball, we need  precise  capability to extrapolate hadron yields to 
unobserved kinematic domains and particle types. This is achieved by 
introducing statistical occupancy  parameters $\gamma_i>1, i=q,s$.
Within  this approach there is good systematic
behavior of physical observables as a function of collision conditions such as energy 
or centrality~\cite{Torrieri:2000xi,Rafelski:2004dp,%
Letessier:2005kc,Letessier:2005qe,Rafelski:2009jr,Kuznetsova:2009wd}.
The yields of hadrons are  in general found not to be in chemical equilibrium, $\gamma_i\ne 1$;
the hadronization temperature is found near to $T\simeq 140$~MeV. 

While this value of $T$ could be  further away from  the deconfinement crossover domain,   
this is where chiral symmetry restoration is achieved~\cite{Fodor:2009ax,Gupta:2009tv},
and  QGP is transformed into hadrons. Moreover, in this approach the variation
of the freeze-out temperature with baryochemical potential parallels the slope seen
in the lattice data. Another  important outcome
of this approach is that a fit to data offers a good statistical  significance. 
 Results obtained can be interpreted in terms of
a dynamical picture of nearly chemically 
equilibrated QGP,  decaying into free streaming hadrons.
The high intrinsic QGP entropy content explains why equilibrated QGP turns into  
chemically overpopulated (over-saturated) HG phase space -- the fast breakup of QGP 
means that the emerging hadrons 
do not have opportunity  to re-establish chemical equilibrium in the HG phase.

{\bf c)} {\it  Single Freeze-out} or/and {\it Strangeness Nonequilibrium} model has as
the main objective  statistically significant description of hadron yields achieved
with minimal effort. Only strangeness chemical
non-equilibrium is allowed. This is often enough to produce a decent data fit and  
to assure that all particles can be formed at the same physical 
condition~\cite{Becattini:2000jw,Broniowski:2001uk,Becattini:2003wp,Becattini:2005xt,Becattini:2008yn}.
The main result of this approach is a hadronization temperature near $T\simeq 160$~MeV which
agrees with Hagedorn temperature~\cite{Hagedorn:1967ua,Hagedorn:1980kb}. 

{\bf Particle ratios of interest:}
We must include in our theoretical consideration of multistrange hadron yields 
the contributing yield of decaying hadron resonances. 
Within SHM these individual yields  generally depend on several parameters. 
The phase space occupancy $\gamma_q$ scales particle yields according to
light quark content, and a similar parameter $\gamma_s$ refers to strange 
quark content. Temperature $T$ quantifies the size of accessible phase space. 
Baryo-chemical potential $\mu_{\rm B}$ differentiates baryons from antibaryons and 
strange chemical potential $\mu_{\rm S}$ does the same for strangeness. There is 
also a potential $\mu_{\rm I3}$ related to different number of up and down quarks
which is constrained by proton and neutron asymmetry in colliding nuclei, and the
overall yield is normalized by a volume parameter $V$. 

By considering  the ratio
\begin{equation} 
   \frac{\Xi}{\phi} \equiv \small{\sqrt{\frac{\overline{\Xi}^+\Xi^-}{\phi\phi}}}
        \simeq \gamma_q  f (T), 
\labeq{1}
\end{equation}
we eliminate in good approximation most  of the SHM parameter dependencies since:\\
{\bf \ \phantom{ii}i)} By taking the product of particle 
and antiparticle, we eliminate baryo-chemical potential $\mu_{\rm B}$ 
as well as strange chemical potential $\mu_{\rm S}$. \\
{\bf \ \phantom{i}ii)} We also eliminate the strange quark phase space
occupancy $\gamma_s$, because the strange and anti-strange quark content 
in the numerator and denominator is the same.\\
{\bf \ iii)} The overall normalization   is eliminated by the 
fact that we have the same number of hadrons in the ratio numerator 
and denominator.

The $\Xi/\phi$ ratio   depends on the probability 
of finding a non-strange $d, \bar d$-quark at the formation of
 $\Xi^-(dss)$ and $\overline\Xi^+(\bar d \bar s \bar s)$, respectively. 
This is expressed by the light quark phase space occupancy 
$\gamma_q$. Furthermore, temperature $T$ controls the magnitude of
\begin{equation} 
f(T)\simeq \sum_i\frac{g_i}{3}\small{\left(\frac{m_{\Xi_i}}{m_\phi}\right)^{3/2}}
\large{ e^{\frac{m_\phi-m_{\Xi_i}}{T}}}
\labeq{2}
\end{equation}
the (non-relativistic) phase space ratio  of $\Xi^-$ and $\phi$. 
{ $\Xi(1321)$  is always a decay product of
$\Xi^*(1530)$. Thus aside of the ground state $i=1:\Xi(1321),\, g_1=2$ 
one must include in the sum the  $\Xi^*$(1530), $g_2=4$ resonance.} 
Consideration of this special yield ratio
parallels the earlier effort made to identify   $\gamma_s/\gamma_q$ 
in Ref.\cite{Rafelski:2002ga}.

We  extend our considerations to include single strange  
${\rm K}^+(u\bar s),{\rm K}^-(\bar u  s)$ mesons  and triple strange 
$\Omega^-(sss), \overline\Omega^+(\bar s\bar s\bar s)$ baryons considering 
the ratios: 
\begin{equation}
 {\Xi\over{\rm K}} \equiv\! 
\small{\sqrt{\frac{\overline\Xi^+\Xi^-}{{\rm K}^+{\rm K}^-}}}=\gamma_s f_1(T); 
\quad
 {\Omega\over\phi} = \!\small{\sqrt{\frac{\overline\Omega^+\Omega^-}{\phi\phi}}}=\gamma_s f_2(T).
 \labeq{5}
\end{equation}
Given the quark content, both ${\Xi/{\rm K}}$ and ${\Omega/\phi}$ are proportional 
to  strange quark yield, i.e. the strange quark phase 
space occupancy    $\gamma_s $ and a function $f_i(T)$.


The  arguments leading to \req{1}, \req{5} are strictly valid only in Boltzmann approximation. 
Considering  quantum   statistics, there is some
residual dependence of $f(T)$ on chemical parameters, involving
higher powers of $\gamma_q$ for the ratio $\Xi/\phi$  \req{1},
and higher powers of $\gamma_s$ for  the ratio $\Xi/{\mathrm K}$  \req{5}. 
In order to estimate the magnitude of the quantum statistics effect we calculate the 
actual particle ratios  with SHAREv2~\cite{Torrieri:2004zz}  using both quantum and Boltzmann statistics.
We find that the Boltzmann approximation we used overestimates $\Xi/\phi$ by $0.25\%$, 
which is always negligible. For $\Xi/{\mathrm K}$,  we find that it is overestimated
by Boltzmann approximation by up to $4\%$, the relatively larger effect is due to 
the relatively low mass of the kaon. Since the experimental error is much greater we continue to 
consider the simple theoretical Boltzmann yields. When further below we consider ratios
involving pions, all results are obtained using SHAREv2, which accounts for resonance
decays and all yields can be obtained using quantum statistics.


{\bf Experimental data:}
We consider $4\pi$ data from the CERN-SPS NA49 experiment, and  for
the STAR experiment at RHIC  the   acceptance rapidity interval 
is $|y| < 0.5$; therefore at RHIC we use the 
yield per unit of rapidity $dN/dy$ and omit the differential 
$dy$ when referring to relative yields. For the $\phi$ meson
we consider the recently published data  from 
STAR~\cite{Abelev:2008zk} 
and the updated data from NA49~\cite{Alt:2008iv}. We collected the 
necessary data for $\Xi$ and $\overline\Xi$ baryons
from Refs.~\cite{Alt:2008qm,Adams:2006ke,Speltz:2006}.
 
We do not use NA49 158 GeV results, 
since these  experimental  results do not allow to  interpolate 
the  different centrality bins  used to measure different 
multistrange particle yields. 
We  could not simply combine data from different centrality bins 
seen  the variation of yields with centrality (that is $\gamma_s$).  
The STAR 62 experiment provides data in several centrality bins,
defined as a percentage
of the most central collisions:  data from the most central collisions  
is found in the centrality $0-5\%$ interval
and the most peripheral collision results presented are 
in $70-80\%$ bin. The relation to $N_{\mathrm{part}}$ and/or impact parameter $b$
is discussed in~\cite{Abelev:2008ez}.

We  use recent data for ${\rm K}^\pm$ mesons from STAR experiments at both 
$\sqrt{s_{NN}}=200$ and $62.4\,\rm{GeV}$ from \cite{Abelev:2008ez}.
For the  SPS NA49 data we use yields from~\cite{Alt:2007fe,Afanasiev:2002mx}.

\begin{table}
\caption{\label{tab:params}Fit parameters used to determine particle yields for incompatible centrality bins using $f(N_{\mathrm{part}}) \equiv a\cdot N_{\mathrm{part}}^b + c$ (see text for details).} 
\begin{ruledtabular}
\begin{tabular}{l|c|c|c}
	&	a	&	b	&	c	\\ \hline
$\pi^-$ & $4.179 \times 10^{-1} $ & $1.072$ & $\phantom{-}7.107 \times 10^{-1} $ \\
$\pi^+$ & $4.247 \times 10^{-1} $ & $1.048$ & $\phantom{-}6.422 \times 10^{-1} $ \\
$K^+$ & $5.433 \times 10^{-2} $ & $1.111$ & $-1.014 \times 10^{-1} $ \\
$K^-$ & $4.812 \times 10^{-2} $ & $1.107$ & $-3.859 \times 10^{-2} $ \\
$\Xi^-$ & $1.228 \times 10^{-3} $ & $1.247$ & $-4.678 \times 10^{-3} $ \\
$\overline{\Xi}^+$ & $8.978 \times 10^{-4} $ & $1.221$ & $\phantom{-}9.390 \times 10^{-4} $ \\
$\phi^0$ & $4.162 \times 10^{-3} $ & $1.203$ & $-9.311 \times 10^{-3} $ \\
\end{tabular}
\end{ruledtabular}
\end{table}

\begin{figure}
\centerline{\hspace*{-0.3cm}\includegraphics[width=2.45in,angle=-90]{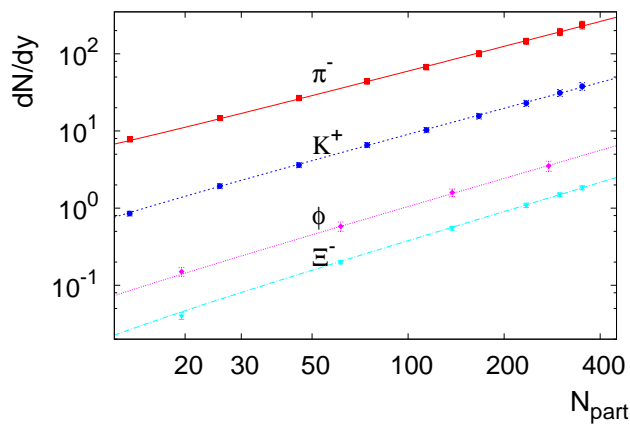}}
\caption{\label{fig:fit} (Color online) 
Data points (full symbols) of particle yields used in the analysis,
and their respective fitted centrality dependence.
}
\end{figure}

We note that different centrality bins are often chosen for different particle types.
In order to  be able to form particle ratios in a common 
centrality interval,  we inter/extrapolate,
that is fit individual  yields as a function of 
the number of participants using a simple functional form
$f(N_{\mathrm{part}}) \equiv a\cdot N_{\mathrm{part}}^b + c$. We show the fit parameters 
$a$,$b$ and $c$  in Table \ref{tab:params} and compare the experimental results and the
fit   in  figure \ref{fig:fit}.

%

\begin{table*}[bth]
\caption{\label{tab:table1}Values of ratios   ${\Xi/\phi}$ \req{1}, ${\Xi/{\rm K}}$ \req{5},  $\Xi/\pi$ and $\phi/\pi$ \req{8} 
obtained from the data and the resulting estimated 
uncertainty in $\gamma_s$ and $\gamma_q$, respectively. When symbol `E' is shown in the error column,
the data ratio is  result of  interpolation and/or extrapolation needed to account for different
centrality bins.}
\begin{ruledtabular}
\begin{tabular}{lcccccccc}
Experiment & Centrality & ${\Xi/\phi}\times 10 $ & $\delta\gamma_q$  & ${\Xi/{\rm K}}\times 10^2$ & $\delta\gamma_s$  &$\Xi/\pi\times 10^3$ & $\phi/{\rm K}\times 10$ & $\phi/\pi\times 10^2$\\ 
\hline
STAR 62    & 0-5\%      & 3.04 & E      & 4.19                 & 9.6\%  &     6.22  & 1.38 & 2.04\\
STAR 62    & 5-10\%     & 3.00 & E      & 4.08                 & 9.2\%  &     6.20  & 1.36 & 2.06\\
STAR 62    & 10-20\%    & 2.94 & E      & 4.06                 & 9.3\%  &     5.98  & 1.38 & 2.04\\
STAR 62    & 20-40\%    & 2.88 & 12.5\% & 3.79 & E      & 5.48 & 1.32 & 1.91 \\
STAR 62    & 40-60\%    & 2.85 & 14.6\% & 3.38 & E      & 4.65 & 1.18 & 1.63 \\
STAR 62    & 60-80\%    & 2.49 & 19.3\% & 2.84 & E      & 3.45 & 1.14 & 1.38\\ \hline
STAR 200   & 0-20\%     & 3.02 & 11.8\% & 4.06                 & 12.9\% & 6.04   & 1.34 & {\footnotemark[1]$2.54^{+0.21}_{-0.09}$} \\ \hline
SPS 80 AGeV& 7\%        & 3.33 & 24.5\% & 3.04                 & 22.7\% & 2.60   & 0.83 & 0.88\\
SPS 40 AGeV& 7\%        & 2.45 & 42.1\% & 1.89                 & 18.0\% & 3.23   & 0.78 & 0.83\\
SPS 30 AGeV& 7\%        & 2.57 & 66.5\% & 1.85                 & 24.3\% & 2.10   & 0.63 & 0.72\\
\end{tabular}
\end{ruledtabular}
\footnotemark[1]{
For STAR 200 $\phi/\pi$ considering  figure 14 in~\cite{Abelev:2008zk} we give an average  
of data  for centralities up to 50\%.}
\end{table*}

\begin{figure}
\centerline{\hspace*{-0.3cm}\includegraphics[width=2.45in,angle=-90]{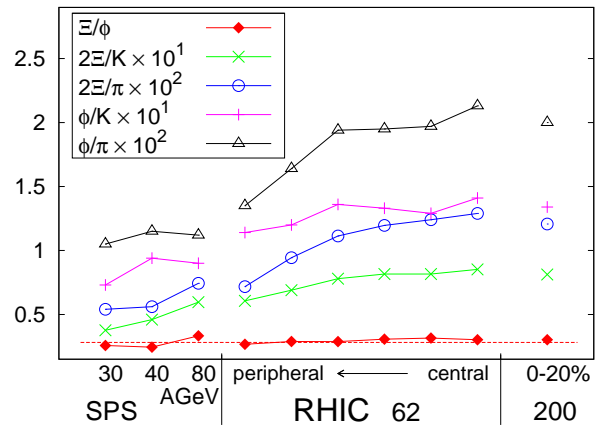}}
\caption{\label{fig:ratios} (Color online) 
Data points of  ${\Xi/\phi}$ \req{1}, ${\Xi/{\rm K}}$ \req{5},  $\Xi/\pi$ and $\phi/\pi$ \req{8}. 
The straight line for  ${\Xi/\phi}=0.281$
 .}
\end{figure}

{\bf Particle ratios:}
After this preparation we can form ratios of particle yields
as shown in figure \ref{fig:ratios} and table \ref{tab:table1}.
We note that the ${\Xi/\phi}$  relative yield  does not change 
much over a wide range of energies and centralities, in contrast to
the individual hadron yields which enter the ratio.
 The average value of all available data points
is ${\Xi/\phi}=0.281$ with an error at 15\% level.

%

\begin{figure}
\centerline{\hspace*{-0.3cm}\includegraphics[width=2.45in,angle=-90]{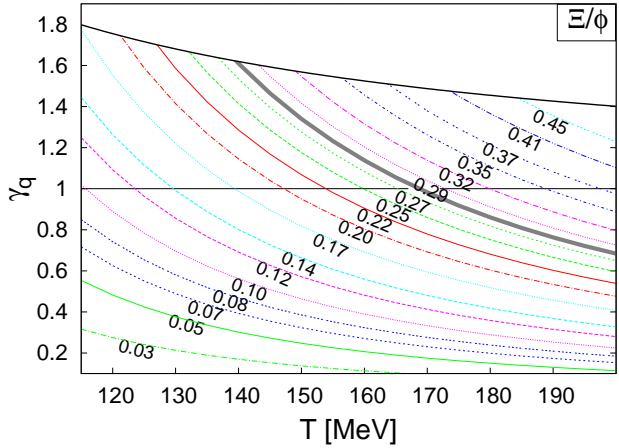}}
\caption{\label{fig:1} (Color online) 
 Lines of a constant given ratio~\req{1} $\Xi/\phi\in [0.03,0.45]$
in the $T,\gamma_q$ plane.
The lines for $\gamma_q = 1$ and $\gamma_q = \gamma_q^{\rm crit}$ 
are presented by solid black lines. The average result, 0.281, of all SPS and RHIC
experiments is highlighted by a thick gray line. As this ratio is considered constant,
this line indicates the prediction of LHC results.}
\end{figure}

The remarkable result, the constancy of ${\Xi/\phi}$  
means that   at SPS and RHIC energies the mechanisms 
and conditions at which double-strange particles
are produced are very similar and that, according to \req{1}, there is a 
constraint between values of $\gamma_q$ and $T$, which we now explore in figure~\ref{fig:1}
where we show in the $T,\gamma_q$ plane 
the  theoretical SHM results as lines  for a constant ratio ${\Xi/\phi}$. These
values are obtained  using SHAREv2 and varying $\gamma_q$ and $T$,
with all other model parameters fixed  to a reasonable physical 
values. In this way we also confirm once again  the analytical formula \req{2}. 

We limit the magnitude of $\gamma_q$ 
by a critical value of light quark phase space occupancy $\gamma_q^{\rm crit}$.
$\gamma_{\pi^0}\equiv \gamma_q^2 \leq (\gamma_q^{\rm crit})^2 
  = \exp({m}_{\pi^0}/T)$, 
which is the condition where the pion phase space distribution function diverges 
for  $\rm{m}_{\pi^0} = 135\,\,\mathrm{MeV/c^2}$. 
The experimental values ${\Xi/\phi}\simeq 0.281\pm$15\% are found consistent
with all SHM models in that for $\gamma_q=1$  we find the value  $T=170\pm 10$ MeV, 
and for $\gamma_q\to 1.63$ a value $T\to 140$ MeV.

{\bf Test of SHM models:}
We have seen that hadronization of $\Xi$ and $\phi$ is consistent
with the three different SHM models, but there is an interesting
constraint between $T,\gamma_q$ 
arising from the constancy of the relative $\Xi$ and $\phi$ yield.
We see also in figure \ref{fig:ratios} that the variation of 
${\Xi/{\rm K}}$ is significant, it changes by a factor of 2.3 in the entire data range.
Considering that we already have established 
by the study of ${\Xi/\phi}$ that the hadronization temperature does not vary 
this indicates that there is 
a  variation of $\gamma_s$ value by a factor of about 2.3 in the data range.
We conclude that a fixed value $\gamma_s=1$ cannot be chosen. This
rules out the SHM model~{\bf a)}. We also note that this argument 
can be made in the same way considering  the variation of the 
other ratios in fig. \ref{fig:ratios}, e.g. $\Xi/\pi$ and $\phi/K$.

\begin{figure} 
\centerline{\hspace*{-0.7cm}\includegraphics[width=5.1in, angle=-90]{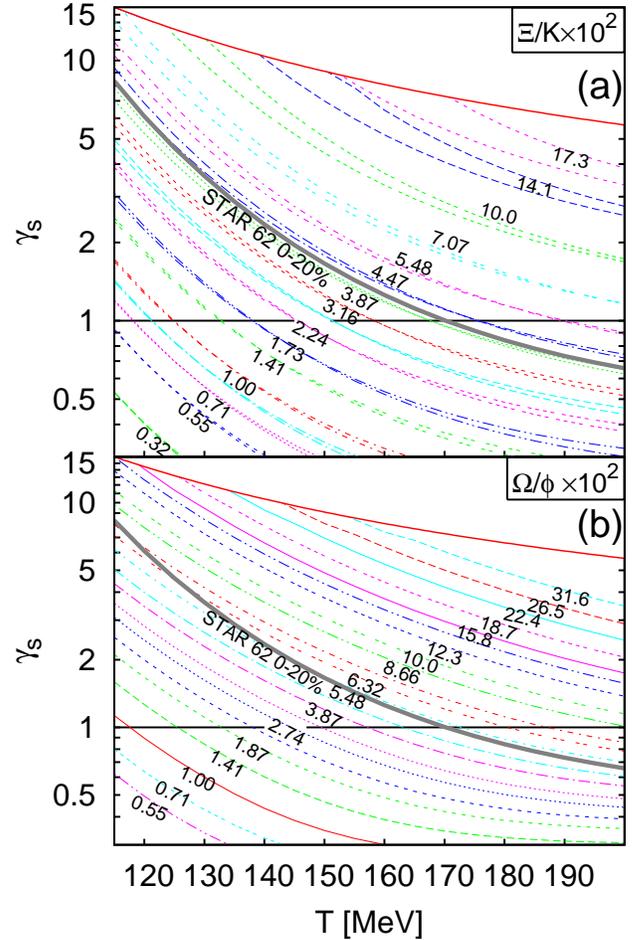}}
\caption{\label{fig:3}(Color online) Lines of constant ratio ${\Xi/{\rm K}}$ (a) 
and ${\Omega/\phi}$ (b).
Experimental data from most central  0-20\% STAR 62 are indicated by a thick  line (b), 
and is assumed in the bottom frame as a prediction. See text for more detail.  }
\end{figure}

SHM results for ${\Xi/{\rm K}}$ and ${\Omega/\phi}$ in $T, \gamma_s$  plane are shown 
in  figure~\ref{fig:3}, obtained by the same method as before; i.e. using SHARE with
 other SHM parameters fixed at an appropriate value.
For a given  ${\Xi/{\rm K}}$ and/or ${\Omega/\phi}$ a slight 
$\gamma_q$ dependence remains, since there are unrelated resonances decaying into $K$ 
(and to lesser degree $\Xi$). Thus we present for each fixed value of $\Xi/K$ 
 two extremes: $\gamma_q=1$, 
and  $\gamma_q = \gamma_q^{\rm crit}$.
The effect is depicted in figure~\ref{fig:3}a  in terms of 
two lines shown by the same line type. 

Note that  similarly as for $\gamma_q$, there is a critical value 
for $\gamma_s$ based  on the Bose-Einstein condensation of 
the $\eta$ meson ($\eta = 0.55(\rm{u\overline u+d\overline d}) + 
0.45\rm{s\overline s}$~\cite{Uvarov:2001wv,Li:2007xf}).
The large values of $\gamma_s$  could be relevant to the future LHC results.
To compare theory and experiment we 
show the thick 0-20\% STAR 62 line and by looking at the bottom frame 
 of figure~\ref{fig:3} we obtain the prediction:
$5.5\times 10^{-2} < {\Omega}/{\phi} < 7.0\times 10^{-2}$, the variation
due to variability of hadronization temperature.

\begin{figure}
\centerline{\hspace*{-0.2cm}\includegraphics[width=2.45in,angle=-90]{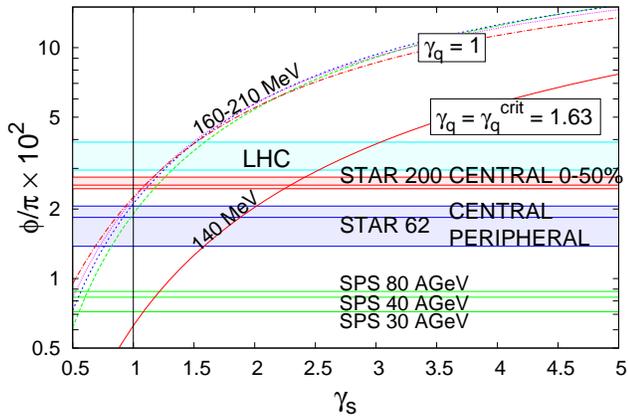}}
\caption{\label{fig:4} (Color online)  
The relative $\phi/\pi$ \req{8} yield as a function of $\gamma_s$ in 
several hadronization scenarios, see text. 
The vertical solid black line shows the chemical 
equilibrium with $\gamma_s = 1$. For 
experimental data see table \ref{tab:table1}. 
Predicted values for LHC are indicated in blue.}
\end{figure}

In order to further elaborate the validity of models~{\bf b), c)}  we 
show  on the right in table~\ref{tab:table1} the ratios $\Xi/\pi$, $\phi/{\mathrm{K}}$,
and $\phi/\pi$, where
\begin{equation}
\frac{\Xi}{\pi}\equiv \sqrt{\frac{\Xi^-\overline \Xi^+}{\pi^-\pi^+}};\quad
\frac{\phi}{\mathrm{K}}\equiv \sqrt{\frac{\phi\phi}{{\mathrm{K}^-}{\mathrm{K}^+}}};\quad 
\frac{\phi}{\pi}\equiv \sqrt{\frac{\phi\phi}{\pi^-\pi^+}}.
 \labeq{8}
\end{equation}
The  experimental $\Xi/\pi$  and $\phi/\pi$  relative yields vary by 
a  factor $\simeq 3.5$ in both cases. In figure~\ref{fig:4}
we show the $\phi/\pi$ ratio and compare to theory as a function of $\gamma_s$ at fixed given $T$.
Model {\bf  b)} with $T\simeq 140, \gamma_q=\gamma_{\rm crit}$ implies that the different
experimental results correspond to  $1<\gamma_s <2.4$. These  values are 
consistent with the large value of $ \gamma_q=\gamma_{\rm crit}\simeq 1.6$.  On the other hand,
for $\gamma_q=1$ several fixed $T$ lines nearly coincide in the
interesting range $210\ge T\ge 160$ MeV. This means that the  growth in yield of $\phi$ is nearly
compensated by the growth in $\pi$-multiplicity. It will be
very interesting to see how LHC results will line up in this presentation,
since we see that the high energy RHIC results even at $\gamma_q=1$  imply  $\gamma_s>1$.
A value $\gamma_s>1$ is incompatible with the picture of strangeness production in
hadron collisions, and implies the presence of a strangeness dense QGP phase as a source of 
hadrons.

{\bf Behavior at LHC:}
As already remarked the ratio  $\Xi/\phi\simeq 0.28$ is firmly constrained and 
cannot change. Even under the extreme LHC conditions we expect that 
this ratio will be the same as at RHIC. However, considerable changes can be
expected for other (multi)strange particle ratios which were discussed 
earlier~\cite{Kuznetsova:2006bh,Letessier:2006wn}. Here we will mainly address
the $\phi/\pi$ ratio. 

When we accept the premise, that entropy and strangeness are conserved 
during the hadronization, we can predict values
of the phase space occupancy $\gamma_s$ in chemical semi- and non-equilibrium models for LHC.
We expect a 20\% increased value of strangeness over entropy $s/S \simeq 0.037$~\cite{Letessier:2006wn}. 
For the two models under consideration
($T=140$ MeV,$\gamma_q=\gamma_q^{critical}$ and $170$ MeV,$\gamma_q=1$) 
this value  suggests~\cite{Kuznetsova:2006bh} $\gamma_s/\gamma_q \simeq 1.55$. 
The expected $\phi/\pi$ ratio is $2.95\times 10^{-2}$ and $3.90\times 10^{-2}$ for the two models 
as indicated by the boundaries of the LHC band in figure~\ref{fig:4}. Experimental results
of this magnitude requires $\gamma_s>1$ and concludes in favor of chemical 
non-equilibrium the still ongoing discussion of chemical equilibrium models.

{\bf Summary and conclusions:}
We find that the relative particle
yield $\Xi/\phi$ is practically constant as function of 
centrality and  reaction energy at RHIC and SPS. We find that
these particles,   despite their small reaction cross-sections 
are emerging at  the  same hadronization condition  
as  all bulk particles. 
This result was anticipated~\cite{Rafelski:2000by}
  for  a fast  expanding QGP fireball which  under-cools   and rapidly
 breaks  apart  (hadronizes), and has been used extensively in single
hadronization models~\cite{Letessier:2005qe,Rafelski:2009jr,Rafelski:2004dp,Broniowski:2001uk} 
 
Variation in  the ratio $\Xi/{\rm K}$ 
(and thus also $\phi/{\rm K}\propto \gamma_s/\gamma_q$) 
implies a variation in strange
phase space occupancy $\gamma_s$, in agreement with the  
expectation that strangeness production grows with energy and 
centrality of the collision. This experimental result is 
incompatible with the chemical equilibrium  model {\bf a)}, 
for which also the parameter $\gamma_s$ is fixed to $1$ by definition.

Considering further  the yields $\Xi/\pi$  and $\phi/\pi$ 
consistency with the bulk matter particle production rates is arrived
 at within the chemical non-equilibrium model 
{\bf b)} with  $\gamma_q>1$ and $\gamma_s>1$.
These values imply   that the observed strange hadrons yields 
are  above   chemical  equilibrium, a feature predicted to be 
 signature for hadronization of a QGP-fireball~\cite{Rafelski:1982ii}.
The expected   further increase of  $\gamma_s>1$ at LHC implies
a further increase of the $\phi/\pi$ ratio, providing   a 
clear distinction between chemical  non-equilibrium model 
{\bf b)} and semi-equilibrium model {\bf c)}.

Our results show that the yields of all multistrange hadrons available today are
1) compatible with the SHM picture of hadron formation,  2) are well described by
current chemical nonequilibrium hadronization models in the parameter domain 
obtained from the other hadron yields, 3) these data are incompatible with  
the chemical equilibrium single-freeze-out SHM. A critical test of our
approach is that in LHC-ion experiments the  $\Xi/\phi$ ratio remains the same
as has been observed at SPS and RHIC.

{\bf Acknowledgments}
JR would like to thank Z.L. Matthews and O. Villalobos Baillie
of Birmingham University and CERN-ALICE
collaboration for interesting  discussions.
This work was supported by a grant from the U.S. Department of Energy, DE-FG02-04ER41318 .


\end{document}